\documentclass[pre,preprint,groupedaddress, showkeys,showpacs]{revtex4}

\usepackage{graphicx}
\usepackage{amsmath}
\usepackage{mathrsfs}

\begin{document}


\title{Rectification and diffusion of self-propelled particles in a two-dimensional corrugated channel}


\author{Bao-quan  Ai $^{1}$} \email[Email: ]{aibq@hotmail.com}
\author{Qiu-yan Chen $^{1}$}
\author{Ya-feng He$^{2}$}
\author{Feng-guo  Li$^{1}$}
 \author{Wei-rong Zhong$^{3}$} \email[Email: ]{wrzhong@jnu.edu.cn}


\affiliation{$^{1}$Laboratory of Quantum Engineering and Quantum Materials, School of Physics and Telecommunication
Engineering, South China Normal University, 510006 Guangzhou, China\\
$^{2}$College of Physics Science and
Technology, Hebei University, 071002 Baoding, China.\\
$^{3}$Department of Physics and Siyuan Laboratory, College of Science and Engineering,
Jinan University, 510632 Guangzhou, China.}


\date{\today}
\begin{abstract}
  \indent Rectification and diffusion of non-interacting self-propelled particles is numerically investigated in a two-dimensional corrugated channel. From numerical simulations, we obtain the average velocity  and the effective diffusion coefficient. It is found that the self-propelled particles can be rectified by the self-propelled velocity. There exist optimal values of the parameters (the self-propelled velocity, the translational diffusion constant, and the height of the potential) at which the average velocity takes its maximal value. There exists an optimal translational diffusion at which the effective diffusion constant is maximal. The self-propelled velocity can strongly increase the effective diffusion, while the large rotational diffusion rate can strongly suppress the effective diffusion.
\end{abstract}

\pacs{05. 60. Cd, 05. 40. -a, 82. 70. Dd}



\maketitle
\section {Introduction}
\indent The problem of rectifying motion in random environments is a longstanding issue, which has many theoretical and practical implications \cite{rmp}. Brownian ratchets have been proposed to model the unidirectional motion driven by zero-mean nonequilibrium fluctuations. Broadly speaking, there are four types of Brownian ratchet models:  rocking ratchets\cite{a1}, flashing ratchets\cite{a2}, correlation ratchets \cite{a3}, and entropic ratchets \cite{a4}. These ratchet models mainly focus on passive Brownian particles. Few work on Browinian ratchet has involved active Brownian particles. As we know, active motion and non-linear dynamics in systems of active objects have received much attention. There are numerous realizations of self-propelled particles\cite{toner,lauga} in nature ranging from bacteria \cite{diluzio,hill,shenoy,leptos} and spermatozoa\cite{riedel} to artificial colloidal microswimmers. The kinetic of self-propelled particles moving in potentials could exhibit peculiar behavior \cite{Schweitzer,Burada,Schimansky-Geier,tailleur,fily,kaiser,bickel,buttinoni,mishra,ohta,peruani,czirok,stark,weber,chen}. Therefore, it is necessary to set up ratchet model in which active Brownian particles can be rectified.

 \indent Rectification of self-propelled particles in asymmetric structures has also attracted much attention. Experimental
studies \cite{Leonardo,Galajda}  show the key role of self-propulsion for rectifying
cell motion for driving a nano-sized ratchet-shaped wheel \cite{Leonardo} or in an array of asymmetric funnels \cite{Galajda}.
Recently, there has been increasing interest in theoretical work on rectification of self-propelled particles \cite{angelani,Wan,ghosh,potosky}.
Angelani and co-workers \cite{angelani} studied the run-and tumble particles in periodic potentials and found that the asymmetric potential produces a net drift speed. Rectification phenomenon of overdamped swimming bacteria was theoretically observed in a system with an array of asymmetric barriers \cite{Wan}.
Ghosh and co-workers \cite{ghosh} studied the transport of Janus particles in periodically compartmentalized channel and found that the rectification can be orders of magnitude stronger than that for ordinary thermal potential ratchets. Potosky and co-workers \cite{potosky} found that even in a symmetric potential a spatially modulated self-propelled velocity can induce the directed transport.

\indent In all these studies, the potential or structure is quasi-one-dimensional and independent in each dimension. In this paper, we extended these studies to the case of a two-dimensional corrugated channel in which the potential is taken as periodic and symmetric in the $x$-direction, whereas in the $y$-direction it works as a trap which strength is periodically modulated in the $x$-direction. Due to the modulation between two directions, a nonzero average drift can be induced even if the potential along $x$-direction is symmetric. The phase shift in the modulation function determines the direction of the transport. The optimal rectification can be obtained by tuning the parameters of the system. The self-propelled velocity can strongly increase the effective diffusion, while the large rotational diffusion rate can strongly suppress the effective diffusion.
\section{Model and methods}

\indent In this paper, we consider self-propelled particles that move in a two-dimensional corrugated channel $U(x,y)=U(x+L,y)$ of period $L$ (shown in Fig. 1).  The potential is periodic in the $x$ direction and parabolic in the $y$ direction. In the overdamped limit, the dynamics of self-propelled particles is described by the following Langevin equations \cite{fily,ghosh}
\begin{equation}\label{fb1}
\frac{dx}{dt}=v_0\cos\theta+\mu F_x+\sqrt{2D_0}\xi_x(t),
\end{equation}

\begin{equation}\label{fb2}
\frac{dy}{dt}=v_0\sin\theta+\mu F_y+\sqrt{2D_0}\xi_y(t),
\end{equation}

\begin{equation}\label{fb3}
\frac{d\theta}{dt}=\sqrt{2D_{\theta}}\xi_\theta(t),
\end{equation}
where $F_x=-\frac{\partial U(x,y)}{\partial x}$ and $F_y=-\frac{\partial U(x,y)}{\partial y}$. $v_0$ is the self-propelled velocity and $\mu$ is the mobility. $\theta$ is the self-propelled angle. $D_0$ is the translational diffusion constant and $D_\theta$ is the rotational diffusion rate, which describes the nonequilibrium angular fluctuation as it may arise, for instance, from tumble dynamics of swimming organisms \cite{berg}. $\xi_{x,y}(t)$ is  the white Gaussian noise and satisfies the following relations,
\begin{eqnarray}
  \langle \xi_i(t)\rangle &=& 0 \\
  \langle \xi_i(t)\xi_j(s)\rangle &=& \delta_{ij}\delta(t-s), \indent i,j=x,y,
\end{eqnarray}
The symbol $\langle...\rangle$ denotes an ensemble average over the
distribution of the random forces. $\delta$ is the
Dirac delta function. $\xi_\theta(t)$ is an additional Gaussian noise which models the fluctuations of the self-propelled angle $\theta$ and it satisfies
\begin{eqnarray}
\langle \xi_\theta(t)\rangle &=& 0 \\
  \langle \xi_\theta(t)\xi_\theta(s)\rangle &=& \delta(t-s).
\end{eqnarray}

 The potential is taken to be a two-dimensional corrugated channel\cite{bao}
\begin{equation}\label{}
  U(x,y)=-U_{0}\sin(\frac{2\pi x}{L})+\frac{1}{2}C_{0}[1-\lambda\sin(\frac{2\pi x}{L}+\phi)]y^{2},
\end{equation}
where $U_0$ is the height of the $x$-direction potential, and $C_0$ is the intensity of the $y$-direction potential. $\phi$ is the phase shift between the $x$-direction potential and the modulation function. $\lambda$ is the modulation constant with $0\leq \lambda <1$.

\indent The classical passive ratchets demand three key ingredients \cite{rmp} (a) nonlinearity: it is necessary since the system
will produce a zero-mean output from a zero-mean input in a linear system; (b)asymmetry
(spatial and/or temporal): it can violate the symmetry of the response; (c) fluctuating input zero-mean force: it can break thermodynamical equilibrium.
Now we will analyse our active system with these ingredients. First, the system is nonlinear due to the nonlinear potential.  We introduce a one-dimensional effective potential through eliminating the $y$-variable: $U_{eff}(x)=-D_0\ln\big[ \int_{-\infty}^{\infty}dy \exp(-\frac{U(x,y)}{D_0}) \big]=U_1(x)+\frac{1}{2}D_0\ln[\frac{C(x)}{2\pi D_0}]$, where $U_1(x)=-U_{0}\sin(\frac{2\pi x}{L})$ and $C(x)=\frac{1}{2}C_{0}[1-\lambda\sin(\frac{2\pi x}{L}+\phi)]$. Although the potential in $x$-direction is symmetric, the effective potential $U_{eff}(x)$  is asymmetric when $\phi\neq 0,\pi, 2\pi$. Finally,
the term $v_0\cos \theta(t)$ in Eq. (\ref{fb1}) is a random force with an exponential correlation function which can break thermodynamical equilibrium. Therefore, our system satisfies the three key ingredients and can show the phenomenon of rectification. In particular, the setup described by Eqs. (\ref{fb1},\ref{fb2},\ref{fb3})  is equivalent to  a pulsating passive ratchet \cite{rmp,Cecchi}, where the term $v_0\cos \theta$ can be seen as the external driving force, the term $\mu F_x$ can be used as a pulsating potential due to the coupling between the precesses $x$ and $y$.

\indent Upon introducing characteristic length scale $L$, time scale $\tau_0$, and energy $U_0$ , Eqs. (\ref{fb1},\ref{fb2},\ref{fb3}) can be rewritten in dimensionless form, namely
\begin{equation}\label{fb11}
\frac{d\hat{x}}{d\hat{t}}=\hat{v}_0\cos\theta+\hat{F}_{\hat{x}}+\sqrt{2\hat{D}_0}\hat{\xi}_{\hat{x}}(\hat{t}),
\end{equation}

\begin{equation}\label{fb21}
\frac{d\hat{y}}{d\hat{t}}=\hat{v}_0\sin\theta+\hat{F}_{\hat{y}}+\sqrt{2\hat{D}_0}\hat{\xi}_{\hat{y}}(\hat{t}),
\end{equation}

\begin{equation}\label{fb31}
\frac{d\theta}{d\hat{t}}=\sqrt{2\hat{D}_{\theta}}\hat{\xi}_{\theta}(\hat{t}),
\end{equation}
with  $\hat{x}=\frac{x}{L}$, $\hat{y}=\frac{y}{L}$,  $\hat{t}=\frac{t}{\tau_0}$, $\hat{U}=\frac{U}{U_0}$, $\tau_0=\frac{L^2}{\mu U_0}$. The remaining rescaled parameters are $\hat{v}_0=\frac{v_0L}{\mu U_0}$, $\hat{D}_0=\frac{D_0}{\mu U_0}$, $\hat{D}_{\theta}=\frac{D_{\theta}L^2}{\mu U_0}$. Then, the potential is rewritten as
\begin{equation}\label{ux}
  \hat{U}(\hat{x},\hat{y})=-\sin(2\pi \hat{x})+\frac{1}{2}\hat{C}_0[1-\lambda\sin(2\pi\hat{x}+\phi)]\hat{y}^2,
\end{equation}
 with $\hat{C}_0=\frac{C_0 L^2}{U_0}$.  From now on, we will use
only the dimensionless variables and shall omit the hat
for all quantities occurring in Eqs. (\ref{fb11},\ref{fb21},\ref{fb31}).
\begin{figure}[htbp]
\begin{center}
\includegraphics[width=9cm]{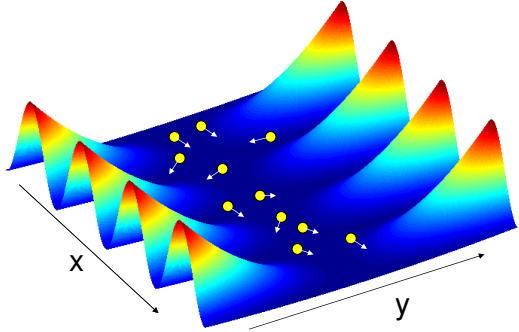}
\vspace{-1.5cm}
\caption{Scheme of the ratchet device: non-interacting self-propelled particles moving in a two-dimensional periodic potential $U(x,y)$, which is periodic in the $x$ direction and parabolic in the $y$ direction. A slice of the potential is a parabola for a given $x$. The particle has the self-propelled velocity with $v_x=v_0\cos\theta$ and $v_y=v_0\sin\theta$.}\label{1}
\end{center}
\end{figure}

\indent Key quantities of particle transport through periodic potentials
are the  particle velocity and the effective diffusion coefficient. Though the movement equation of Brownian particles in
the present system can also be described by the corresponding Fokker-Planck equation \cite{fk}, it is very difficult to obtain the analytical expressions of the average velocity and the effective diffusion coefficient. The behavior of the quantities of interest can
be corroborated by Brownian dynamic simulations performed by integration of the Langevin equations using the second-order stochastic Runge-Kutta algorithm.  Because the potential along the $y$-direction is parabolic, we only calculate the $x$-direction average velocity and effective diffusion coefficient. From Eqs. (\ref{fb1},\ref{fb2} \ref{fb3}), the $x$-direction average velocity can be obtained from the following formula
            \begin{equation}\label{V}
            v=\lim_{t\rightarrow\infty}\frac{\langle x(t)\rangle}{t},
            \end{equation}
and the average probability current $J=v/L$.

 \indent The $x$-direction diffusion coefficient $D_{x}$ can be calculated by the formula
 \begin{equation}\label{Dx}
            D_x=\lim_{t\rightarrow\infty}\frac{1}{2t}\langle[x(t)-\langle x(t) \rangle]^2 \rangle.
\end{equation}

\section{Results and Discussion}
\indent For the numerical simulations, the total integration time was more than $5\times 10^5$ and the transient effects were estimated and subtracted. The integration step time $\Delta t$ was chosen to be smaller than $10^{-4}$.  The stochastic averages reported above were obtained as ensemble averages over $3 \times 10^{4}$ trajectories with random initial conditions. With these parameters, the simulation results are robust and do not depend on the time step, the integration time, and the number of trajectories.  It will be shown that the average velocity , the effective
diffusion coefficient are some functions of the various parameters in the system.

\subsection{Directed transport}
\begin{figure}[htbp]
\begin{center}
\includegraphics[width=8cm]{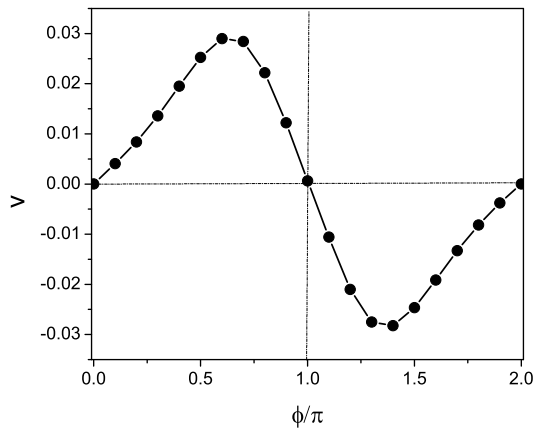}
\caption{Average velocity $v$ as a function of the phase shift $\phi$. The other parameters are $C_0=5.0$, $\lambda=0.9$, $v_0=2.0$, $D_0=0.6$, and $D_\theta=0.03$. }\label{1}
\end{center}
\end{figure}
\indent The average velocity $v$ as a function of the phase shift $\phi$ is reported in Fig. 2. It is found that the sign of $v$ is determined by the phase shift $\phi$. The average velocity is positive for $0<\phi<\pi$, zero at $\phi=0$, $\pi$, and $2\pi$, negative for $\pi<\phi<2\pi$. Therefore, we can have current reversals by changing the phase shift. In addition, there exists an optimal value of $\phi$ at which the magnitude of $v$ takes its maximal value.
\begin{figure}[htbp]
\begin{center}
\includegraphics[width=8cm]{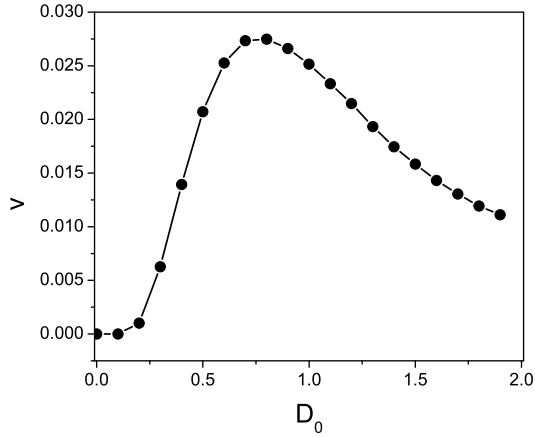}
\caption{Average velocity $v$ as a function of the translational diffusion constant $D_0$. The other parameters are $C_0=5.0$, $\lambda=0.9$, $v_0=2.0$, $\phi=0.5\pi$, and $D_\theta=0.03$.}\label{1}
\end{center}
\end{figure}

\indent Figure 3 shows the average velocity $v$ versus the translational diffusion constant $D_0$. The curve is observed to
be bell shaped, which shows the feature of resonance. When $D_0\rightarrow 0$, the particle cannot pass the barrier and stays at the bottom of the potential, so the average velocity $v$ tends to zero.  When $D_0\rightarrow \infty$ such that the thermal noise is very large, the ratchet effect disappears and the average velocity $v$ goes to zero, also.  Therefore, there is an optimal value of $D_0$ at which the average velocity $v$ takes its maximal value.

\begin{figure}[htbp]
\begin{center}
\includegraphics[width=8cm]{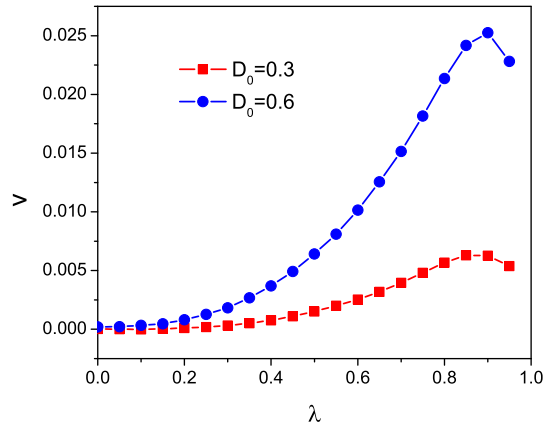}
\caption{Average velocity $v$ as a function of the modulation constant $\lambda$ for different values of $D_0$. The other parameters are  $C_0=5.0$, $v_0=2.0$, $\phi=0.5\pi$, and $D_\theta=0.03$.}\label{1}
\end{center}
\end{figure}
\indent  Average velocity $v$ as a function of the modulation constant $\lambda$  is shown in Fig. 4 for different values of $D_0$.  When $\lambda=0$, there is no coupling between the $x$-direction potential and the $y$-direction potential, the total potential is always symmetric, so the ratchet effect disappears.  As the modulation constant $\lambda $ increases, the average velocity first increases, and then decreases near $\lambda=0.9$. There exists an
optimal value of $\lambda$ at which the average velocity is maximal.

\begin{figure}[htbp]
\begin{center}
\includegraphics[width=8cm]{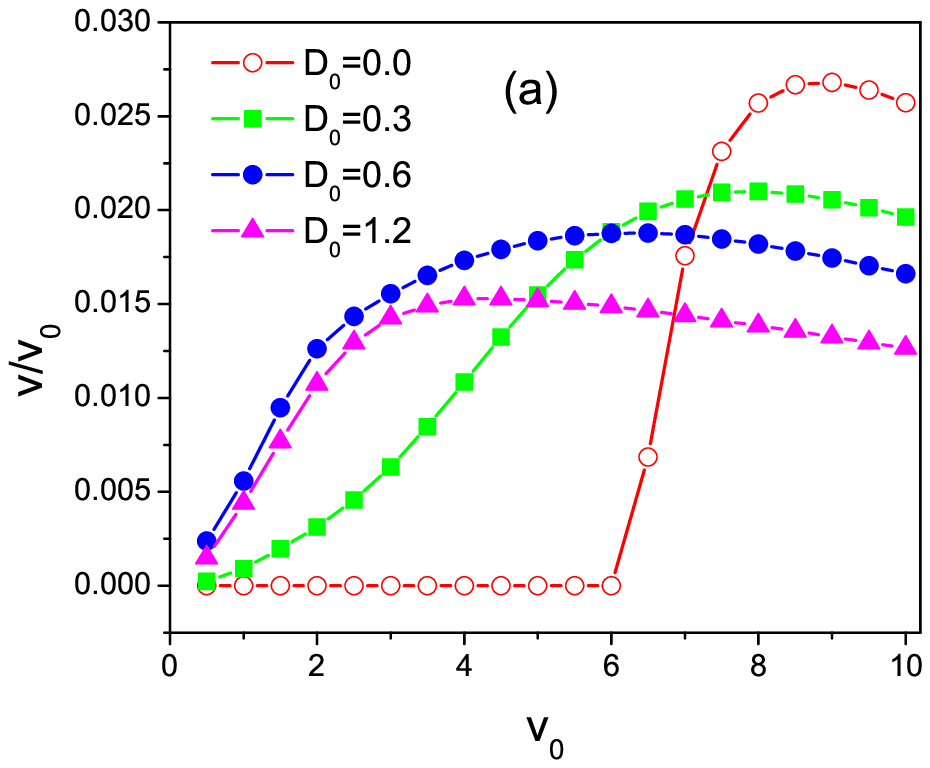}
\includegraphics[width=8cm]{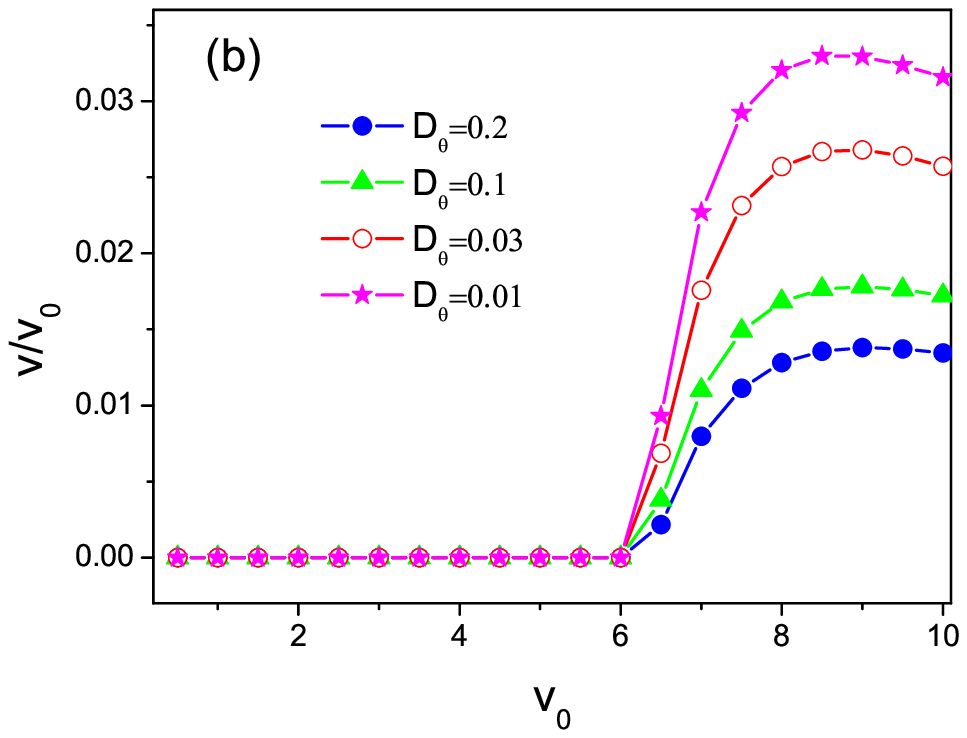}
\caption{The ratio $v/v_0$ as a function of the self-propelled velocity $v_0$. (a)For different values of $D_0$ at $D_\theta=0.03$. (b) For different values of $D_\theta$ at $D_0=0.0$. The other parameters are $\lambda=0.9$, $C_0=5.0$, and $\phi=0.5\pi$.}\label{1}
\end{center}
\end{figure}

\indent In Fig. 5 (a),  the ratio  $v/v_0$  is depicted as a function of the self-propelled velocity $v_0$ for several values of $D_0$.  The curves are observed to be bell shaped, there exists an optimal value of $v_0$ at which the ratio $v/v_0$ takes its maximal value.  As the translational diffusion constant $D_0$ increases, the position of the peak shifts to the small values of $v_0$.  Therefore, the optimal self-propelled velocity can facilitate the rectification of the particles. Figure 5 (b) shows the ratio $v/v_0$ versus $v_0$ for zero translational diffusion ($D_0=0$). It is found that the average velocity is zero until $v_0$ reaches a critical value $F_{c}=2\pi$. From the potential shown in Eq. (\ref{ux}), the maximal force for the potential at $y=0$ is $2\pi$.  Therefore, in order to pass through the channel, the active particle needs a nonzero $v_0$ to move against this force $F_{c}$. For the case of the finite $D_{0}$, the noise helps the particle to cross potential barriers and thereby activates a nonzero current even for $v_0<2\pi$ (shown in Fig. 5(a)).

\begin{figure}[htbp]
\begin{center}
\includegraphics[width=8cm]{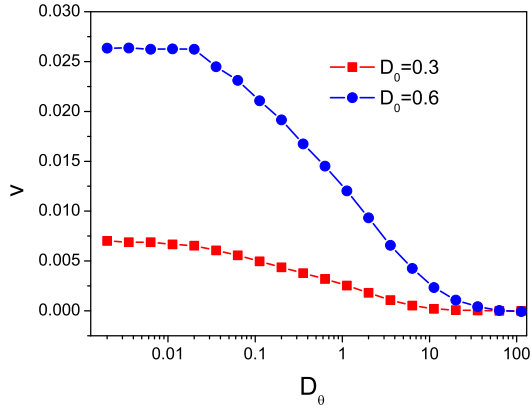}
\caption{Average velocity $v$ as a function of the rotational diffusion rate $D_\theta$ for different values of $D_0$. The other parameters are $\lambda=0.9$, $C_0=5.0$, $v_0=2.0$, and $\phi=0.5\pi$.}\label{1}
\end{center}
\end{figure}

\indent Figure 6 depicts the average velocity $v$ as a function of the rotational diffusion rate $D_\theta$. When $D_\theta\rightarrow 0$, the self-propelled angle $\theta$ almost does not change, the average velocity approaches its maximal value,  which is similar to the adiabatic case in the forced thermal ratchet \cite{a1}.  As $D_\theta$ increases, the average velocity $v$ decreases. When $D_\theta\rightarrow\infty$, the self-propelled angle changes very fast, the particle was trapped in the valley of the potential, so the average velocity tends to zero, which is similar to the high frequency driving case in the forced thermal ratchet \cite{ai}.
\begin{figure}[htbp]
\begin{center}
\includegraphics[width=8cm]{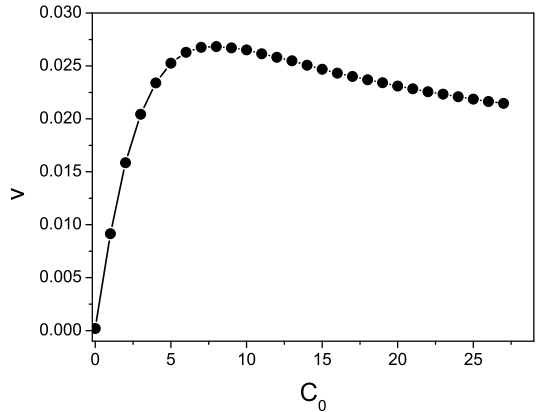}
\caption{Average velocity $v$ as a function of the constant $C_0$. The other parameters are $\lambda=0.9$, $v_0=2.0$, $D_0=0.6$, $\phi=0.5\pi$, and $D_\theta=0.03$.}\label{1}
\end{center}
\end{figure}

\indent In Fig. 7, we plot the average velocity $v$ versus the intensity $C_0$ of the $y$-direction potential. When $C_0\rightarrow 0$, $U(x,y)$ reduces to $U_1(x)$, the potential is completely symmetric, so the average velocity goes to zero.  When $C_0\rightarrow \infty$, the potential $U_1(x)$ can be neglected, the effect of the phase shift disappears, the average velocity also tends to zero. Therefore, the optimal intensity $C_0$ can facilitate the rectification of self-propelled particles.

\subsection{Effective diffusion coefficient}

\begin{figure}[htbp]
\begin{center}
\includegraphics[width=8cm]{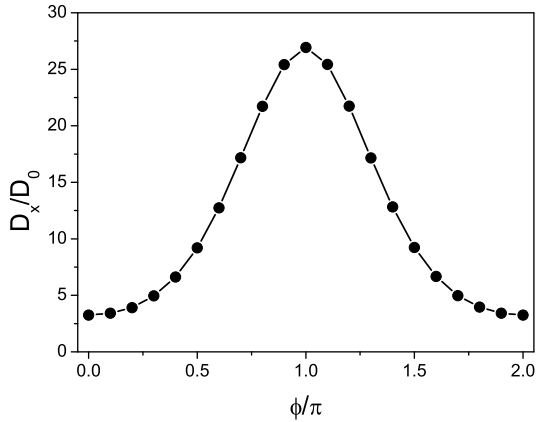}
\caption{The $x$-direction diffusion coefficient $D_{x}/D_0$ versus the phase shift $\phi$. The other parameters are $C_0=5.0$, $\lambda=0.9$, $v_0=2.0$, $D_0=1.0$, and $D_\theta=0.03$. }\label{1}
\end{center}
\end{figure}

\indent Figure 8 shows the $x$-direction diffusion coefficient $D_{x}/D_0$ versus the phase shift $\phi$. It is found that the curve is symmetric with respect to $\phi=\pi$. There is a  peak in the curve and the $x$-direction diffusion coefficient $D_{x}/D_0$ takes its maximal value at $\phi=\pi$. The effective diffusion coefficient strongly depends on the phase shift.

\begin{figure}[htbp]
\begin{center}
\includegraphics[width=8cm]{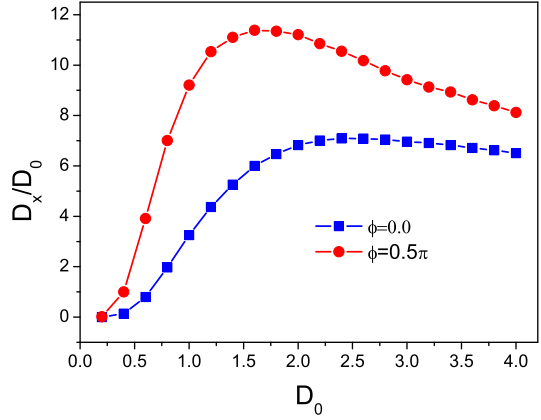}
\caption{The $x$-direction diffusion coefficient $D_{x}/D_0$ versus the translational diffusion constant $D_0$ for different values of the phase shift $\phi$. The other parameters are $C_0=5.0$, $\lambda=0.9$, $v_0=2.0$, and $D_\theta=0.03$. }\label{1}
\end{center}
\end{figure}

\indent  The $x$-direction diffusion coefficient $D_{x}/D_0$ as a function of the translational diffusion constant $D_0$ is displaced in Fig. 9 for both rectification ($\phi=0.5\pi$)  and non-rectification ($\phi=0$) cases.  When $D_0\rightarrow 0$, the particle cannot pass across the $x$-direction barriers and stays at the bottom of the potential, so the effective diffusion tends to zero. When $D_0\rightarrow\infty$, $D_0$ dominates the diffusion, the $x$-direction diffusion coefficient $D_{x}/D_0$ tends to $1$.  There exists an optimal value of $D_0$ at which the diffusion coefficient takes its maximal value.  In addition, the scaled effective diffusion coefficient for the rectification case is always larger than that for the non-rectification case, which shows that the ratchet effect facilitates the diffusion of self-propelled particles.
\begin{figure}[htbp]
\begin{center}
\includegraphics[width=8cm]{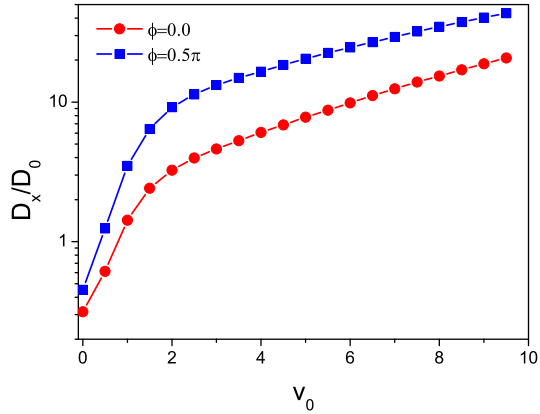}
\caption{The $x$-direction diffusion coefficient $D_{x}/D_0$ versus the self-propelled velocity $v_0$  for different values of the phase shift $\phi$. The other parameters are $C_0=5.0$, $D_0=1.0$, $\lambda=0.9$, and $D_\theta=0.03$. }\label{1}
\end{center}
\end{figure}

\indent  Figure 10 depicts  $x$-direction diffusion coefficient $D_{x}/D_0$ versus the self-propelled velocity $v_0$ for both rectification ($\phi=0.5\pi$) and non-rectification ($\phi=0$) cases. It is found that the $x$-direction diffusion coefficient $D_{x}/D_0$ increases monotonically with the self-propelled velocity.  Therefore, the self-propelled velocity can strongly facilitate the diffusion of self-propelled particles.

\begin{figure}[htbp]
\begin{center}
\includegraphics[width=8cm]{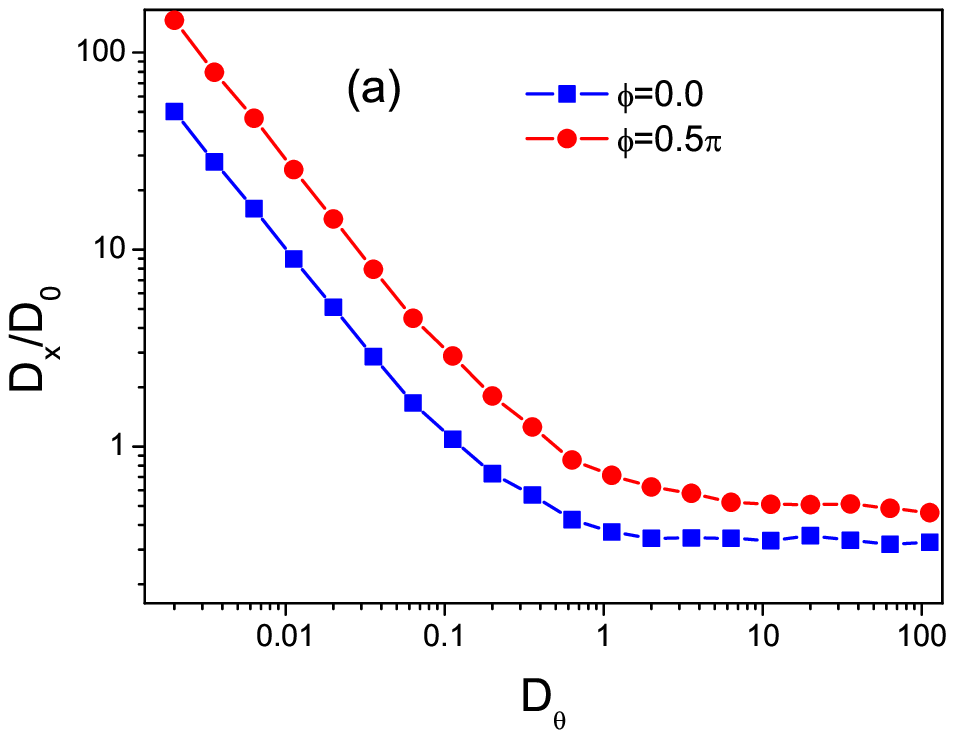}
\includegraphics[width=8cm]{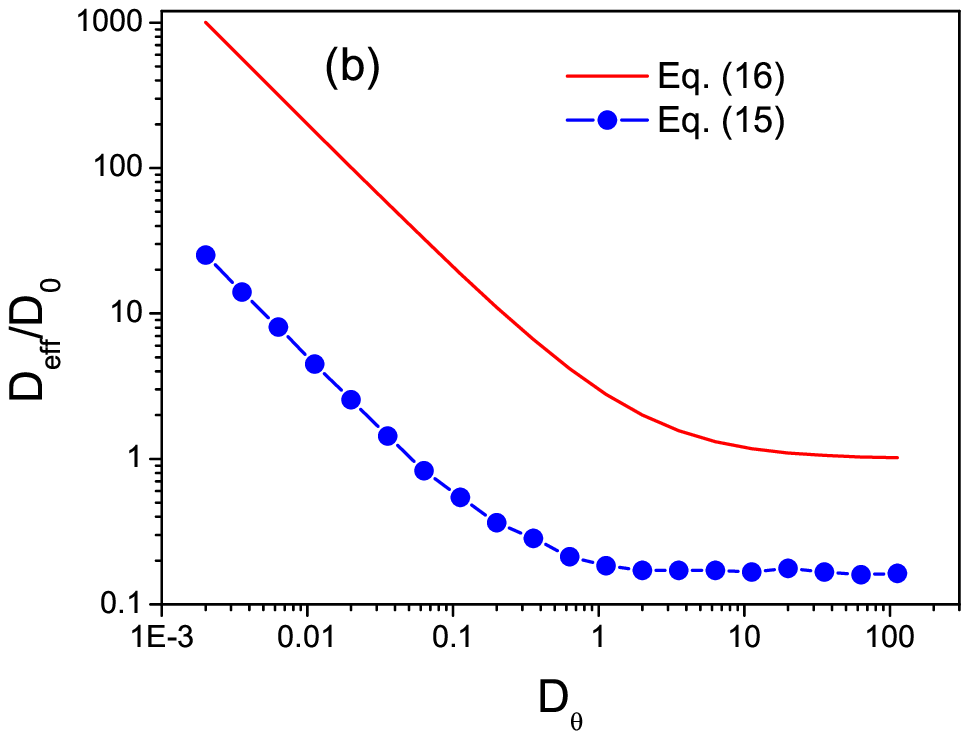}
\caption{(a) The $x$-direction diffusion coefficient $D_{x}/D_0$ versus the rotational diffusion rate $D_\theta$ for different values of the phase shift $\phi$. (b)The two-dimensional effective diffusion coefficient $D_{eff}/D_0$ versus the rotational diffusion rate $D_\theta$ for with and without potentials.  The other parameters are $C_0=5.0$, $D_0=1.0$, $v_0=2.0$, and $\lambda=0.9$ }\label{1}
\end{center}
\end{figure}

\indent Figure 11 (a) shows the $x$-direction diffusion coefficient $D_{x}/D_0$ as a function of the rotational diffusion rate $D_\theta$ for both rectification ($\phi=0.5\pi$)  and non-rectification ($\phi=0$) cases. As the rotational diffusion rate increases, the $x$-direction diffusion coefficient $D_{x}/D_0$
monotonically decreases. When $D_\theta\rightarrow\infty$, the self-propelled angle changes very fast, the particle is trapped in the valley of the potential. Therefore, the large rotational diffusion rate can strongly suppress the diffusion of of the self-propelled particles.

\indent In Figs. (8-11), we mainly focus on the $x$-direction diffusion coefficient. However, the present setup is a two-dimensional system, it is necessary to investigate the total effective diffusion coefficient.  The two-dimensional effective diffusion coefficient $D_{eff}$ without drift term  can be calculated from the mean-squared displacement \cite{weber,Reimann}
\begin{equation}\label{deff}
  D_{eff}=\lim_{t\rightarrow\infty}\frac{\langle |\vec{r}(t)-\vec{r}(0)|^2 \rangle}{4t},
\end{equation}
where $\vec{r}(t)=\{x(t),y(t)\}$  and $\vec{r}(0)=\{x(0),y(0)\}$ are the finial and initial position vectors. $\langle |\vec{r}(t)-\vec{r}(0)|^2 \rangle$ is the mean-squared displacement. In the absence of the potential, there is a well-known analytical expression for the effective diffusion coefficient \cite{howse}
\begin{equation}\label{deff2}
  D_{eff}=D_0+\frac{1}{4}v_0^{2} \tau_{\theta},
\end{equation}
with $ \tau_{\theta}=2/D_{\theta}$. The two-dimensional effective diffusion coefficient $D_{eff}/D_0$ as a function of the rotational diffusion rate is shown in Fig. 11(b). From Fig. 11(a) and (b), we can find that the effective diffusion coefficient from two dimensions is similar to that from one dimension ($x$-direction). In addition, we can also find that the effective diffusion coefficient without potentials is much larger than that with potentials.

\section {Concluding remarks}
\indent In this paper, we numerically studied the transport of self-propelled particles in a two-dimensional corrugated channel. It is found that a nonzero average drift can be induced by the self-propelled velocity. The
direction of the transport is determined by the phase shift between the x-direction potential and the
coupling function. The average velocity is positive for $0<\phi<\pi$, zero at $\phi=0$, $\pi$, and $2\pi$, negative for $\pi<\phi<2\pi$. There exist optimal parameters (e. g. the self-propelled velocity, the translational diffusion constant, the modulation constant, the intensity of the $y$-direction potential)  at which the average velocity takes its maximal value.
The average velocity decreases with increasing the rotational diffusion rate. The effective diffusion coefficient is strongly affected by the phase shift and  takes the maximal values near at $\phi=0.5\pi$ and $1.5\pi$.  The scaled effective diffusion coefficient for the rectification case is always larger than that for the non-rectification case. There exists an optimal value of the translational diffusion constant at which the scaled effective diffusion coefficient is maximal. The self-propelled velocity can strongly facilitate the effective diffusion, while the large rotational diffusion rate can strongly destroy the effective diffusion.

\indent  This work was supported in part by the National Natural Science Foundation
of China (Grant Nos. 11175067, 11004082, and 11205044), the PCSIRT (Grant No. IRT1243), the Natural
Science Foundation of Guangdong Province, China (Grant No. S2011010003323).

\end{document}